\documentstyle[12pt,epsf]{article}
\newlength{\overeqskip}
\newlength{\undereqskip}
\newcommand{\nc}{\newcommand}
\nc{\be}{\begin{equation}} 
\nc{\ee}{\end{equation}}
\nc{\bea}{\begin{eqnarray}}
\nc{\eea}{\end{eqnarray}}
\nc{\bi}[1]{\bibitem{#1}}
\nc{\sss}{\scriptscriptstyle}
\nc{\lsim}{\mbox{\raisebox{-.6ex}{~$\stackrel{<}{\sim}$~}}}
\nc{\gsim}{\mbox{\raisebox{-.6ex}{~$\stackrel{>}{\sim}$~}}}
\nc{\nn}{\nonumber}
\begin{document}
%
% -----------------------------------------------------------------
% Title Page
% -----------------------------------------------------------------
%
\begin{titlepage}
\pagestyle{empty}
\baselineskip=21pt
\rightline{Report No. LPT Orsay 00-25,UNIL-IPT/00-05}
\vskip 0.8in

\begin{center} 
  {\LARGE\bf Baryogenesis from `electrogenesis' \\
in a scalar field dominated epoch}
\end{center}
\vskip .2in

\begin{center}
     {\Large Michael Joyce}\\
\vskip 0.1 in
  {\it LPT, Universit\'e Paris-XI, B\^atiment 211,
F-91405 Orsay Cedex, France\\
   E-mail: Michael.Joyce@th.u-psud.fr}  \\
\end{center}
%\vskip 0.1 in

\begin{center}
     {\Large Tomislav Prokopec}\\
\vskip 0.1 in
      {\it Universit\'e de Lausanne, Institut de Physique Th\'eorique,
BSP, CH-1015 Lausanne, Suisse}\\
{\it E-mail: Tomislav.Prokopec@ipt.unil.ch}
\end{center}
\vskip 0.2 in

\centerline{ {\bf Abstract} }
\baselineskip=18pt
\vskip 0.3truecm
\noindent
Scalar fields can play a dominant role in the dynamics of the Universe 
until shortly before nucleosynthesis. Examples 
are provided by domination by a kinetic mode of a scalar field, 
which may be both the inflaton and the late time `quintessence',
and also by more conventional models of reheating.
The resultant modification to the pre-nucleosynthesis expansion rate 
can allow solely an asymmetry in right handed electrons to produce 
a net baryon asymmetry when reprocessed by the anomalous $B+L$ 
violating processes of the standard model. The production of such a 
source asymmetry - what we term `electrogenesis' - requires
no additional $B$ or $L$ violation beyond that in the standard
model. We consider a specific model for its generation, by a
simple perturbative out-of-equilibrium decay of Higgs like scalar 
fields with CP-violating Yukawa couplings to the standard 
model leptons. We show that, because of the much enhanced expansion
rate, such a mechanism can easily produce an adequate asymmetry
from scalars with masses as low as $1$TeV. Kinetic
mode domination is strongly favoured because it evades large
entropy release which dilutes the asymmetry.
We also discuss briefly the effect of the abelian hypercharge anomaly.

\end{titlepage}

\baselineskip=20pt

%
%%%%%%%%%%%%%%%%%%%%%%%%%%%%%%%%%%%%%%%%%%%%%%%%%%%%%%%%%%%%%%%%%%%%%%%%%%%%%%%
%  MAIN TEXT
%%%%%%%%%%%%%%%%%%%%%%%%%%%%%%%%%%%%%%%%%%%%%%%%%%%%%%%%%%%%%%%%%%%%%%%%%%%%%%%
%
\section{Introduction}

Until a few years ago cosmology with scalar fields was almost synonymous
with cosmological inflation. Recently there has been an enormous upsurge
in interest in the possibility that scalar fields can play an important
role in the dynamics of the Universe at recent epochs, mainly due to
the observations of the apparent magnitudes of distant supernovae 
\cite{Supernovae} which may be explained by the presence of
such a component \cite{quintessence}. In this context it is 
certainly interesting to consider what the role of such fields can be 
at other epochs, and in particular how their behaviour between the 
end of inflation and their reappearance today might influence 
cosmology in the intervening period.
This question is also related to the `fine-tuning' problem associated 
with such scenarios: how is it that such a field can give a significant
contribution to the energy density today starting from a natural set of 
initial conditions after inflation? This apparent problem is in fact
resolved in a wide class of potentials \cite{jf,tracking}
which generically have the property that in some part of the 
potential they may support modes which are dominated by the 
kinetic energy of the scalar field, so that their
energy density scales away faster than that in the radiation, 
{\it i.e.} $\rho_\phi\propto a^{-n}$, with $4<n\leq 6$, 
where $a$ is the scale factor. 
In principle there is no reason why such modes cannot initially dominate
over the radiation component, and in certain specific models this
is realized.  The main important observational constraint is that such 
domination must terminate by the nucleosynthesis epoch, when the
expansion law must be that given by radiation domination with the
standard model degrees of freedom. There may be an additional contribution
which, conservatively, must be less than about $20 \%$ of the total
\cite{nucleosynthesis}. 

More generally we can consider the question of the cosmology
of the Universe between the end of an inflationary epoch and 
the entry into radiation domination before nucleosynthesis. For a
transition from a scalar field dominated cosmology to occur the
energy in the scalar field must either decay (directly or indirectly) 
into standard model particles - as in standard reheating 
scenarios \cite{kolbturner} -
{\it {or}} it must red-shift away more rapidly than the radiation.
Or some combination of the two can occur.
In the former case any scaling less rapid than that 
during inflation ($n>2$, or equivalently an equation of state 
$p_\phi=w_\phi\rho_\phi$, with $w_\phi>-1/3$) can be
envisaged, with the case $n=3$ corresponding to the most standard 
reheating during the oscillation of the inflaton about the minimum
of a quadratic potential. There is a continual release of entropy
until the radiation dominated epoch, leading to a dilution of
most relevant physical quantities sourced during the scalar
dominated phase. In the latter case, which corresponds to domination 
by the kinetic energy of a homogeneous scalar field (or
equivalently to an equation of state $p_\phi=w_\phi\rho_\phi$, with 
$w_\phi>1/3$ ) the scalar field simply redshifts away until 
it becomes the sub-dominant component.
There is no entropy release, and correspondingly
a coherent energy remains in the scalar field which, given an
appropriate potential (the `self-tuning' potentials of \cite{jf},
or the `tracking' potentials of \cite{tracking}) can become relevant 
again at late times \cite{spokoiny,jp,peeblesvilenkin}. 

In \cite{mj,jp} we have considered in a generic way the effect
of a change in the expansion rate prior to nucleosynthesis on models
of electroweak baryogenesis\footnote{The effects on 
dark matter freeze-out can be inferred from
the work of \cite{Barrow, KamTurner}, who studied mainly modifications 
associated with anisotropy in the expansion.}, 
in particular on the effect on the 
sphaleron bound and the `no-go' theorem for electroweak baryogenesis
in the case of a second order phase transition. As concrete realizations
of such cosmologies we considered models which go through an epoch
after inflation - which, following \cite{mj} we termed `kination' - 
of domination 
by a kinetic mode of a scalar field. This
occurs most naturally in a model in which the universe `reheats' not
by the decay of the inflaton, but by gravitational particle creation
at the end of the inflationary epoch \cite{ford, spokoiny}. In a recent paper 
\cite{dlr} it has been observed that, for low (sub-electroweak) 
reheat temperatures in more traditional models of reheating - in which 
the inflaton decays while oscillating in a mode with matter scaling 
after inflation - the effects discussed in  \cite{mj,jp} on electroweak 
cosmology also result. There is in this case an even larger relative 
boost to the expansion rate (see below), but
a very large entropy release which tends to undo any of the enhancing
effects of the greater expansion rate. In \cite{tp} one of us (TP)
has considered the general case of a decaying inflaton evolving in
a mode scaling as $1/a^n$, and shown that, while the same larger 
boost to the expansion rate occurs as in the $n=3$ case of \cite{dlr},
the entropy release problem is greatly reduced as the kinetic mode 
$n=6$ limit is attained.

Here we concentrate on another aspect of such alternative cosmologies,
which is a simple consequence of the observation which has 
been made in \cite{js,jp,dlr}: Because of the enhanced expansion
rate, the right-handed electrons of the standard model may remain 
out of equilibrium until a temperature below the electroweak 
phase transition. It is well known that asymmetry in right-handed 
electrons - because of their late equilibration 
time -  may be important from  at least two points of view:

$\bullet$ Since right-handed electrons couple to other particles
in the standard model with only an extremely small Yukawa coupling, 
they remain out of equilibrium in an expanding Universe until
relatively late - in the standard radiation dominated cosmology 
until $T \gsim 20$TeV \cite{cdeo, cko}. A pre-existing baryon asymmetry 
can survive the effect of standard model anomalous processes 
- which violate $B+L$ and are unsuppressed until the electroweak phase
transition - only if there are non-zero CP-odd conserved global charges 
when they are operative.
In the absence of such charges the equilibrium attained will 
be CP invariant with zero baryon number. 
As noted in \cite{cko} above \footnote{The scale quoted in 
\cite{cko} is $10$GeV. The increase by a factor 2 is due to a tighter 
bound on the Higgs mass.} $20$TeV  right electron number 
$e_R$ is in fact such an
effective charge, and as a result other global charges like $B-L$
%in which an asymmetry can be stored 
can be violated until close to this scale. 
This leads \cite{cko} to a very significant reduction in the 
bounds on $B-L$  violating interactions in grand unified theories with the
structure appropriate for them to generate baryon asymmetry.
Here the consequences are much simpler and more dramatic: If the
$e_R$ remain out of equilibrium all the way until the electroweak
scale, a baryon number will result from this due to the 
$B+L$ violating processes. When the electroweak scale is reached 
this baryon number will simply be frozen when the $B+L$ violating
processes abruptly switch off. This will be the case irrespective 
of whether there is primordial $B$ or $L$ (or $B-L$), and irrespective 
of whether these charges are violated or conserved. Just like in the 
case of electroweak baryogenesis all the non-trivial physics required 
is in principle present in the standard model. The problem of
baryogenesis then becomes posed as what we will refer to as
`electrogenesis', the generation of the source right handed 
electrons prior to the time at which the $B+L$ violating
processes become suppressed. It is this process which we discuss below.

$\bullet$ The effective conservation of $e_R$ in the early Universe
due to the fact that its perturbative decay channel is out of
equilibrium is not exact, because the $e_R$ charge
has an axial anomaly under the U(1) of hypercharge. There are no
degenerate vacua as in the non-abelian case, but there are finite
energy modes of the U(1) field with Chern-Simons number which can
`eat' the charge. In fact, as discussed in~\cite{js, giovshap} this leads
to an instability at finite density to the formation of long wavelength 
modes of hypermagnetic field. When these modes come inside the horizon they
can evolve during the time in which the right electron number is without its
perturbative decay channel. Here this scenario will be modified as a result 
of the change in the expansion rate, since the perturbative channel
does not come into play until the electroweak scale, at which time
a first order phase transition may produce the turbulence needed to
amplify the produced seed magnetic fields.

\section{Scalar fields and the expansion rate after inflation}

The inflationary solutions for scalar fields represent
only one part of a much wider range of possible behaviours of the energy
density in the zero modes of scalar fields. The full range
can be characterized by the equation of state for a (real) scalar field 
which is determined by the relative weight of the kinetic and
potential energy (see \cite{mj,jp} for a discussion):   
\be
p_\phi=w_\phi\rho_\phi, \quad 
w_\phi=\frac{\frac{1}{2}\dot{\phi}^2 - V(\phi)}{\frac{1}{2}\dot{\phi}^2
+ V(\phi)}, \qquad \rho_\phi \propto a^{-3(w_\phi+1)} .
\label{kin.eR.1}
\ee
The limit of potential energy domination gives inflation, with
$\rho_\phi \approx const$, while the opposite limit of complete
kinetic energy domination gives the most rapid possible red-shifting
of the energy to be $\rho_\phi \propto 1/a^6$. While inflation is
associated with flat potentials (satisfying `slow-roll' conditions),
the latter limit is associated with steep potentials\footnote{The
exception is a flat direction with no associated potential energy
e.g. a Goldstone direction associated with a broken exact global
symmetry, which only has pure kinetic modes.}. A particularly
useful `yard-stick' of flatness/steepness is the simple exponential
potential 
\be
V_{\rm exp}(\phi) = M_P^4 e^{-\lambda\phi/M_P}
\label{kin.eR.2}
\ee
where $M_P=1/\sqrt{8 \pi G} \approx 2.4 \times 10^{18}$GeV 
is the reduced Planck mass (and the origin of $\phi$ has been
chosen to give the simple normalization). This potential in fact 
has an attractor solution \cite{halliwellbarrow} for 
any $\lambda^2<6$ in which the energy
density scales as $1/a^{\lambda^2}$, and as $1/a^6$ for $\lambda^2>6$. 
A potential with a varying slope, {\it e.g.} the inverse power-law potential
of \cite{ratrapeeb, steinhardt} 
$V \sim M_P^{4+\alpha}/\phi^\alpha$ then supports
a kinetic mode at small $\phi$, but an inflationary type (or
`quintessence') mode at large values of the field. Alternatively
an oscillating mode about the minimum of a potential 
$V\sim \lambda_\alpha\phi^\alpha$ gives a broad range of scalings with 
$\rho_\phi \propto a^{-6\alpha/(\alpha+2)}$ \cite{turner}, 
producing thus the familiar
matter scaling when $\alpha=2$ and radiation scaling when $\alpha=4$.

In \cite{jp} we discussed several ways in which a period of kinetic
mode domination (which, following \cite{mj}, we termed 
`kination') could come about after inflation\footnote
{Such kinetic modes have also recently been used to propose a solution 
to the  cosmological moduli problem \cite{dine}.}.
We considered only the case in which the relevant field (the `kinaton')
did not decay itself, and discussed two possible sources for the radiation
in the Universe: The entropy associated with particle creation during
the de Sitter phase (see below), or a more conventional source 
in the decay of a distinct inflaton field. In the latter case 
specific conditions need to be satisfied by the `kinaton' field
to allow it to dominate over the energy produced by the inflaton,
whereas in the former the kinaton and the inflaton are one field
and the domination by the kinetic mode for a period is a 
built-in and necessary feature. 

Our concern in this paper is not the inflationary model building
aspect of the problem, but rather the problem of `electrogenesis'
in this kind of cosmology, as well as in the more conventional reheating
models discussed in \cite{dlr,tp}.  
For the sake of clarity and simplicity we limit ourselves here 
to two definite and simple models with scalar field dominance
continuing until temperatures just above the nucleosynthesis
scale, exemplifying these two types of different cases:

\begin{figure}[t]
\centering
\hspace*{-2mm}
\leavevmode\epsfxsize=12cm \epsfbox{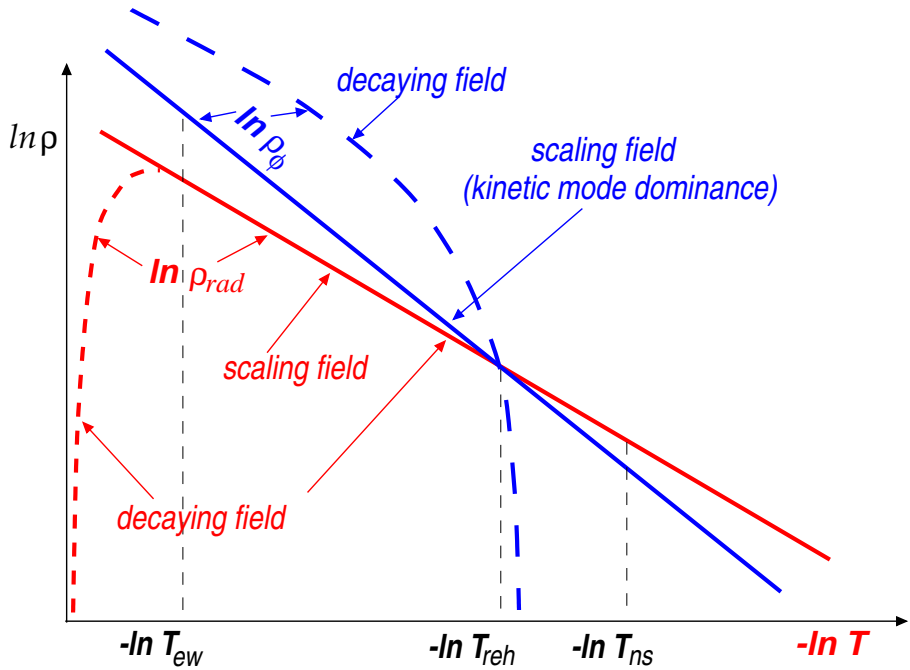}\\[3mm]
\caption[fig2]{\label{fig2} Evolution of energy density in radiation and 
the dominant scalar field as a function of temperature. 
Two cases are illustrated: 
{\it Model (A)} in which the dominant scalar component scales faster
then radiation, but does not decay ({\it solid lines}), and
{\it Model (B)} in which the scalar field decays ({\it dashed lines}).}
\end{figure}

$\bullet$ {\it Model (A):\/} The inflaton rolls after inflation into
a steep potential in which the field rolls in a kinetic mode, so 
that the energy density scales as $1/a^n$ where $n>4$ 
(see Figure~\ref{fig2}). The field is assumed to be very
weakly coupled and the only radiation present is the very sub-dominant
component due to particle creation at the end of the preceeding 
de Sitter phase. The latter has a characteristic energy density $H_I^4$, 
where $H_I$ is the expansion rate at the end of inflation, so that 
initially $\rho_{\rm rad}/\rho_\phi \sim (H_I/M_P)^2$. Provided the
inflaton scales faster than radiation it will become subdominant
at a subsequent time and the transition to radiation domination
is achieved without any decay of the field \cite{ford, spokoiny}.
Requiring that this transition occurs before nucleosynthesis
gives an absolute lower bound on $H_I$, which for 
a pure (or almost) kinetic mode scaling as $1/a^6$ results 
is $H_I \gsim 10^7$GeV (see \cite{jp}). As noted by 
Spokoiny \cite{spokoiny}, for an appropriate potential the field 
can again dominate in a slowly scaling mode at late times. 
This kind of model has been dubbed `quintessential inflation' and 
studied in more detail in \cite{peeblesvilenkin, giovan} 
(see also \cite{rosati}).  

Taking the reheat temperature $T_{\rm reh}$ to be defined 
\footnote{Note that in these models the Universe is strictly
speaking not `reheated' at all - the entropy is left behind
at the end of the de Sitter phase and the important process is
the red-shifting away of the dominant energy in the inflaton. 
Here we adopt the standard definition of `reheat temperature'
as used in standard reheating models. In \cite{jp} we 
used `reheat temperature' to mean the temperature of the radiation 
when it first thermalizes, which is far higher ($\sim 0.1 H_I$) than 
the `reheat temperature' as defined here. What we now call $T_{\rm reh}$
is denoted  $T_{k, end}$ (`end of kination') in \cite{jp}.}
as that when $\rho_\phi\simeq \rho_{\rm rad}$, it is easy to infer  
\footnote{For simplicity we neglect here and elsewhere the small 
reheating factors associated with particle decouplings.}
that above this temperature we have
\begin{equation}
 H = H_{\rm rad}\left(\frac{T}{T_{\rm reh}}\right)^\frac{n-4}{2} \,,
\label{kin.eR.6}
\end{equation}
where $H_{\rm rad}\simeq 1.4\times 10^{-16}(T^2/100{\rm GeV})$
is the standard radiation dominated evolution of the expansion
rate, and $n$ gives the scaling of the energy density in the dominant
scalar mode $\rho_\phi \propto 1/a^n$, with clearly the largest 
enhancement of the expansion rate for the limit $n=6$.
The constraint that the energy density in the scalar field
be less than about $20 \%$ at nucleosynthesis requires 
that $T_{\rm reh} \gsim 5^{1/(n-4)} T_{\rm ns}$ (and $T_{\rm ns}=1$MeV). 
Here we are interested in the case when right electrons are
out of equilibrium at the electroweak scale, which corresponds
therefore to the upper bound 
\begin{equation}
T_{\rm reh}< T_{\rm ew} \left(\frac{H_{\rm ew}}{\Gamma_{e_R}}
     \right)^{\frac{2}{n-4}} ,
\label{kin.eR.7}
\end{equation}
which for the optimum case ($n=6$) becomes 
\begin{equation}
T_{\rm reh}< T_{\rm ew} \frac{H_{\rm ew}}{\Gamma_{e_R}}
   \sim \frac{T_{\rm ew}}{200} \sim 0.5 \,{\rm GeV},
\label{kin.eR.8}
\end{equation}
where we have made use of the fact that the interaction
rate for right electrons through their Yukawa coupling
is $\Gamma_{e_R}\approx 10^{-13}x_{e_R}^2T$ \cite{cko}, and  
we took $x_{e_R}\equiv m_H/2T_{\rm ew}\sim 0.5$ correspondinding
to the current lower bound on the Higgs mass $m_H$. In terms
of the expansion rate the bound (\ref{kin.eR.8}) corresponds
to the requirement of a boost by about $200$ times in the expansion rate
at the electroweak scale relative to that in the standard radiation
dominated cosmology.

$\bullet$ {\it Model (B):\/} The inflaton evolves after inflation into
a potential, in which it rolls or oscillates, scaling as
$1/a^n$ with $6\geq n \geq 3$. The dominant source of entropy
comes from the decay of the inflaton, which is however sufficiently
weakly coupled that reheating occurs between the electroweak scale
and the nucleosynthesis scale. The energy density-temperature dependence 
for this case is illustrated in Figure~\ref{fig2}. The phase we are 
discussing corresponds to the `preheating' phase \cite{preheating}
of inflationary models with the usual mechanism of reheating from 
inflaton decay, either in an oscillatory mode 
(with $n=3$ for a $\phi^2$ potential) or
a rolling mode. We assume here for simplicity perturbative reheating, 
but note that the nonperturbative decay channels of narrow resonance 
may also be considered. A realization of the latter with a rolling
mode is given by the `NO' models of \cite{felderkofmanlinde}.

It is quite easy to show \cite{tp} that in {\it Case (B)\/}
the expansion rate as a function of the temperature is independent
of the equation of state~(\ref{kin.eR.1}), {\it i.e.} the following 
{\it universality in scaling} in the expansion rate holds
\begin{equation}
 H = \frac{5-3w_\phi}{6}\frac{\rho_r}{\Gamma_\phi M_P}
    = H_{\rm rad}\left(\frac{T}{T_{\rm reh}}\right)^2 \,,
\label{kin.eR.4}
\end{equation}
which implies that, for a reheat temperature 
below the electroweak transition, the expansion 
rate is enhanced by $(T/T_{\rm reh})^2$ with respect to the standard rate
$H_{\rm ew}\equiv H_{\rm rad}(T_{\rm ew})$. The condition that 
the right electrons remain out of equilibrium until the electroweak scale
is in this case
\begin{equation}
T_{\rm reh}< T_{\rm ew} \left(\frac{H_{\rm ew}}{\Gamma_{e_R}}
     \right)^{\frac{1}{2}} \sim \frac{T_{\rm ew}}{15} \sim 5\, {\rm GeV}
\label{kin.eR.5}
\end{equation}
which again corresponds to the same minimal boost in the expansion
rate by a factor of about $200$. The extra increase in the expansion rate
as a function of temperature compared to the first case is due to
the `leaking' of the scalar field energy into the radiation. Note that
the lower bound on $T_{\rm reh}$ is in this case $T_{\rm reh} \gsim 2$MeV.

Our interest here finally is in the ratio of baryon number to entropy,
and so will need to include the dilution effect of this entropy production
subsequent to the scale $T_{\rm dec}$ at which the baryon number, or in
fact the source for it, right electron number is produced. 
As discussed in \cite{tp} the entropy
per comoving volume $S_{\rm com}$ scales as $a^3T^3 \propto T^{-3(8-n)/n}$
since $a\propto t^{\,2/n} \propto T^{\,-8/n}$. Thus the dilution factor
$f_{\rm dil}$ due to entropy production between the two scales  
is 
\begin{equation}
f_{\rm dil}\simeq 
\left(\frac{T_{\rm dec}}{T_{\rm reh}}\right)^{\frac{3(8-n)}{n}}\,. 
\label{kin.eR.9}
\end{equation}
Thus there is a very significant difference between the case of the matter
scaling (considered in \cite{dlr}) 
giving $f_{\rm dil}\simeq (T_{\rm dec}/T_{\rm reh})^5$ and that of
the kinetic mode limit with $f_{\rm dil}\simeq T_{\rm dec}/T_{\rm reh}$. 
The origin of this difference can be easily understood: a scalar kinetic mode
gets rid of most of its energy by the rapid red-shifting.

We now turn to the effect of these modifications to the
pre-nucleosynthesis expansion rate on the generation of a baryon
asymmetry from an $e_R$ asymmetry.

\section{From $e_R$ to a baryon asymmetry}

Before discussing the generation of right electron asymmetry in
these cosmologies, we discuss the conversion of such an
asymmetry to a baryon asymmetry when $B+L$ violating processes
are active. `Conversion' is in fact a little misleading as these
processes of course only act on the left-handed fermions: As will
become more explicit now the physics of the creation of the baryon
asymmetry is that the right electrons carry the gauge charge 
hypercharge, which is globally zero and exactly conserved.
When there is net hypercharge in the $e_R$ sector, there must
be also a compensating hypercharge in the rest of the particles.
When this is non-zero the $B+L$ violating processes minimize the
free energy with a non-zero baryon number.   

We follow a standard procedure and consider the equilibrium 
abundance of baryon number subject to the constraints imposed by 
the charges conserved by the fast interactions 
(which are in equilibrium at that time). Because baryon number
violation freezes out at the electroweak scale, when the sphaleron
processes become suppressed, this is the scale at which we need
to calculate baryon number. Above the electroweak scale the rate of
the B-violating processes is mediated by the symmetric phase 
sphaleron transitions, which are unsuppressed: 
$\Gamma_{\rm sph} \approx 25\alpha_w^5 T \sim 10^{-6} T$ \cite{gm},
so that they will have time to  equilibrate above the 
electroweak scale in the models we are discussing ({\it cf.} 
Eqs.~(\ref{kin.eR.6}) and~(\ref{kin.eR.4})). In fact we shall
assume for simplicity that the expansion rate is such that the
$e_R$ are the only standard model degrees of freedom out of equilibrium.
Within the scenarios we are discussing this is not necessarily the 
case, as the rate could be enhanced in principle enough to take also
other heavier particles out of equilibrium. For example the
$\mu_R$ has a Yukawa coupling larger by about $10^2$ and therefore a decay
rate faster by a factor of $(y_\mu/y_e)^2\sim 10^4$, while with a reheat
temperature sufficiently close to the nucleosynthesis scale the
expansion rate may be boosted in {\it Model A\/} by almost as much as
$T_{\rm ew}/T_{\rm ns}\sim10^5$, and in {\it Model B} by 
$(T_{\rm ew}/T_{\rm ns})^2\sim 10^{10}$ times (with of course the
correspondingly large entropy release factor).

With the following hierarchy of couplings
\begin{equation}
\Gamma_{e_R}\ll H\ll \Gamma_{\rm sph}, \Gamma_{\mu_R},..
\end{equation}
the appropriate equilibrium calculation of baryon number is particularly
simple. In the standard model the only conserved charges are the 
gauge charges, $e_R$ and $\frac{1}{3}B-L_i$, where the latter 
is the baryon minus lepton number in each generation. We will make
the slight simplification of assuming only total $B-L$ as conserved
(which would be appropriate at this scale in certain models 
including neutrino mass matrices), which leads to minor numerical 
changes to the results quoted here. (We refer the reader 
to \cite{jp} where the full set of constraint equations can be found.)

To arrive at the set of constraint equations one expresses
the charge densities in terms of particle densities $n_\alpha$ 
using $n_\alpha-\bar n_\alpha =(T^2/6)k_\alpha\mu_\alpha$, 
where\footnote{We use 
here the massless approximation, to which there will be small corrections
due to thermal masses. Note that we also assume the right electron 
distributions can be described by a chemical potential, which is
justified given their relatively fast elastic scattering rate 
through weak hypercharge processes $\sim 10^{-2}T$ \cite{jpt}.} 
$k_\alpha=1 (2)$ for fermions (bosons) and $\mu_\alpha$ is 
the chemical potential for a species $\alpha$. 
Further $\mu_\alpha$ can be re-expressed in terms of the 
chemical potentials for charges $Q_A$ as follows:  
$\mu_\alpha=\sum_A q^A_\alpha\mu_A$, where 
$q^A_\alpha$ is the $A$-charge of the $\alpha$ species. With this we 
can have ({\it cf.} \cite{jp}) the following constraint equations
while the baryon number is out of equilibrium:
\begin{eqnarray}
 Y &=&\frac{T^2}{6}\left[(10+n)\mu_Y+8\mu_{B-L}+\mu_{e_R}\right]
\nonumber\\
 B-L &=&\frac{T^2}{6}\left[8\mu_Y+13\mu_{B-L}-\mu_{e_R}\right]
\nonumber\\
 e_R &=&\frac{T^2}{6}\left[-\mu_Y-\mu_{B-L}+\mu_{e_R}\right] .
\label{conserved-charges}
\end{eqnarray}
Here we used the hypercharge assignments such that 
$Q=Y+T^3$, where $Q$ denotes the electric charge and $T^3$ the isospin.
We have not written the second linearly independent gauge charge 
explicitly, as choosing it as $T^3$ it is simply proportional to
its own chemical potential, and so trivially drops out of the 
equations when we impose $T^3=0$. 
The baryon number $B$ can itself be expressed in terms of
the relevant chemical potentials as
\begin{equation}
 B =\frac{T^2}{6}\left[2\mu_Y+4\mu_{B-L}\right] .
\label{B-number}
\end{equation}
The gauge charge $Y$ must be zero, and
then given any value of the global conserved charges
Eqs.~(\ref{conserved-charges}) can be solved to give the 
baryon number~(\ref{B-number}). 
When $B-L$ is conserved we thus have
\begin{eqnarray}
 B = \frac{2(9+2n)}{59+12n}\;e_R + \frac{2(11+2n)}{59+12n}\,(B-L) 
\label{eR.B1}
\end{eqnarray}
and see that $e_R$ is an almost equally strong source for baryon number
as is $B-L$. Indeed, as $n$ changes from $n=1$ to $n=\infty$, the coefficient
of $e_R$ changes from $0.31$ to $1/3$, while that of $B-L$ from $0.32$
to $1/3$. Hence Eq.~(\ref{eR.B1}) may be quite well approximated by 
\begin{equation}
 B \approx \frac{1}{3}\,\left[e_R + (B-L)\right]  .
\label{eR.B2} 
\end{equation}
Thus, if $B-L$ is conserved by all interactions after inflation,
it is zero and remains zero, but the final baryon number, in 
contrast to the usual radiation dominated universe, is now non-zero
and simply proportional to the original $e_R$ asymmetry. Indeed
therefore we see explicitly that no $B$ violation other than 
that of the anomalous processes of the standard model is 
required to produce it. 

While the latter is the case which will interest us, it is interesting
to note that the result that one obtains a non-zero baryon number 
from $e_R$ is very robust, and is relatively insensitive to whether
the other charges are violated. Indeed it is easy to see that if
$e_R$ is the {\it{only}} conserved charge -- $B-L$ may for example be
violated by some interactions all the way down to the electroweak scale
-- the net baryon number is still non-zero. Indeed, solving the reduced set of
constraint equations for $Y$ and $e_R$ only, with $\mu_{B-L}$
set to zero, we find
\begin{equation}
 B = -\frac{2}{11+n}\,e_R  
\label{B-L violated.}
\end{equation}
which is slightly smaller and of the opposite sign than 
the result in Eq.~(\ref{eR.B1}). 

We conclude that, irrespective of constraints on the value 
of $B-L$ and assumption on whether $B-L$ is conserved, a right-handed
electron asymmetry is reprocessed into a baryon asymmetry of the same order.

\section{Electrogenesis}

We now consider explicitly {\it models for electrogenesis} -- production of a
right-handed electron asymmetry -- prior to the electroweak scale. In the 
standard radiation dominated cosmology right electrons have been understood 
to be of interest because of their capacity to protect a baryon asymmetry
from erasure \cite{cdeo,cko}. Thus their generation has been considered in the
context of theories which also produce such a primordial baryon
or lepton asymmetry, and thus typically the scale characterizing
their generation is very high, around the GUT scale or in the case
of leptogenesis as low as $10^{10}$GeV \cite{leptogenesis}. 
In the present context right electrons are in their own right 
adequate sources for baryogenesis by reprocessing with 
standard model $B+L$ violation. Given that the physics required 
to generate them is CP-violating only, and thus potentially can 
be associated in simple ways with  much lower energy scales, it
is certainly of interest to consider mechanisms which can produce
them quite independently of $B$ or $L$ violation beyond the
standard model.

In fact in the cosmologies being considered one is forced to
seek such different mechanisms of $e_R$ generation for
another very simple reason which we have not drawn attention to so far:
The maximum temperature $T_{\rm max}$ attainable after inflation in 
these cosmologies is in fact much lower 
than in the standard radiation dominated cosmology. Given the requirements
of $T_{\rm reh}$ in (\ref{kin.eR.7}) and (\ref{kin.eR.5}), we can bound 
the temperature  above by extrapolating the expansion rate to the point
$H \sim 0.1 T$. For model A this corresponds to $T_{\rm max} \lsim 10^8$GeV,
while for model B it gives $T_{\rm max} \lsim 10^6$GeV. Above this point 
thermodynamic temperature can have no meaning as the age of 
the Universe is shorter than the equilibration time of any 
process. Thus any mechanism which in the ordinary radiation 
dominated scenario relies on temperatures being reached 
higher than this is not applicable, and we must seek mechanisms
which operate at a lower temperature. 

Here our aim is not to be exhaustive about possible mechanisms,
but rather to study an explicit model which produces an 
$e_R$ asymmetry sufficiently large to source the observed baryon 
asymmetry in these cosmologies. Given that in principle all the
elements are present in the standard model itself, it is natural to
ask -- just as one does in the context of electroweak baryogenesis --
whether it alone might suffice. While in the standard radiation
dominated cosmology the standard model has apparently insurmountable
problems on two fronts \cite{shaprub-review} -- the sphaleron bound 
and the inadequacy of standard model CP violation -- here the former
does not provide a significant constraint. All we require here is that
the right electron number come into equilibrium after the $B+L$ violation
goes out of equilibrium. This is in contrast with baryogenesis scenarios
at a first order electroweak phase transition in which the sphaleron rate
at the transition is required to drop below the expansion rate. So could the 
standard model with its CP violation produce the $e_R$ asymmetry? 
Given that its production can only come about through the same
Yukawa coupling channel, the answer would seem to be definitively
in the negative. In general however the question in these cosmologies
can be framed more generally given that the expansion rate can change
enormously: Is it possible to generate some CP-odd charge (not
necessarily $e_R$) which is conserved on a time scale longer 
than that associated with the $B+L$ violating processes in 
the unbroken phase? We will return briefly to this question
in the conclusion.

Here, just as one does in the context of baryogenesis models, we
add some extra CP-violating physics in the scalar sector.  
We study a simple out-of-equilibrium decay of scalar particles
with CP-violating decays. Interestingly we find that, again because
of the modified expansion rate prior to nucleosynthesis, the mass
of these scalars need not be so far above the electroweak scale
for the mechanism to work. This suggests that the kind of
mechanism for `electrogenesis' we discuss may be 
implementated successfully in other theories with additional 
scalars particles at scales not far above the electroweak scale, with
signatures testable at accelerators.  We will return to this 
point in our conclusions.

\subsection{The Model}
The additional particle content we assume over the standard model
(and the inflaton) is a set of Higgs-like scalar doublets $\Phi^a$ 
coupled to the standard model leptons through a Yukawa type interaction,
{\it i.e.} with interaction Lagrangian
\begin{equation}
{\cal L}_{add}  = - h^a_{ij}\Phi^a \bar\psi_{iL}\psi_{jR} + h.c.,
\end{equation}
where the couplings  $h^a_{ij}$ are CP-violating, {\it i.e.} 
${\bf h^a}^\dag\neq {\bf h^a}$, where ${\bf h^a}$ is the matrix of couplings.
While in principle
CP violation does not mandate a matrix of couplings, but only
a coupling to the right electron itself with a complex phase
unremovable by phase transformations on the whole Lagrangian, 
we will require the flavour mixing structure and
the existence of at least two such scalars in order to implement
the generation of a CP-violating asymmetry. The strongest 
constraints on the masses and the couplings of such scalars 
come from the fact that they are flavour changing. For leptons
the strongest constraint of this type comes from the bounds on the
decay $\mu \rightarrow e \gamma$ \cite{sheryuan, sher}. For couplings
$h$ of order one this requires masses $M_\Phi \gsim 100$TeV, with
the branching ratio for this process going parametrically as
$h_{\mu \tau}^2h_{e \tau}^2(M_W/M_{\Phi})^4$ so that much smaller
masses can be permitted if the couplings have a 
hierarchy like that in the standard model 
Yukawa couplings \cite{sher, hall}. 

\subsection{The Out-of-Equilibrium Conditions}

We consider here a simple out-of-equilibrium decay scenario
for these particles, very analogous to that which occurs
in standard GUT scale baryogenesis scenarios \cite{kolbturner}.
It is possible that nonperturbative decay mechanisms may be operative 
and work just as well, but we limit our treatment here to the simpler 
perturbative case. The perturbative decay rate for $\Phi$ can 
be well approximated by  
\begin{equation}
\Gamma_{\phi,\,\rm pert} = \frac{\vert{\bf h}\vert^2}{8\pi} E_\phi ,
\label{Phi decay rate pert}
\end{equation}
where $E_\phi$ is the energy of $\Phi$, 
$\vert{\bf h}\vert^2=\rm{Tr}({\bf h}{\bf h}^\dag)$  and we have 
assumed the energy of
the $\Phi$ is much greater than that of the produced fermions
({\it e.g.} in the case $m_i=m_j=m$ there is a simple suppression
$E_\phi\rightarrow [E_\phi^2-4m_{\psi}^2]^{1/2}$). 

Before considering the production of a CP asymmetry we first discuss
the out-of-equilibrium condition.  When the particles 
decay, with rate given by (\ref{Phi decay rate pert}), the reverse 
process (or any other one) creating them must be suppressed. This 
is fulfilled if the temperature of the plasma at the time of decay 
is well below the mass scale of the scalars, {\it i.e.} 
\begin{equation}
M_\Phi>T,  \qquad {\rm when}\quad \Gamma_\phi \sim H .
\label{mass phi}
\end{equation}
Equations~(\ref{kin.eR.6}) and~(\ref{kin.eR.4}) give the boost  
to the expansion rate with respect to the radiation dominated 
case as $(T/T_{\rm reh})^p$, where $p=1$ for kinetic mode 
domination ($n=6$), and $p=2$ for a decaying dominant component. 
Making use of this and Eq.~(\ref{Phi decay rate pert}), we infer that 
the constraint~(\ref{mass phi}) can be re-expressed as 
\begin{equation}
M_\Phi> T_{\rm dec} > (70 g_*)^{-1/2(1+p)}\; 
\left[{\vert{\bf  h}\vert}^2 M_PT_{\rm reh}^p\right]^{\frac{1}{1+p}}
        \qquad (0\leq p\leq 2) ,
\label{mass phi constr.}
\end{equation}
where $T_{\rm dec}$ is the temperature at which $\Phi$ decays, and
we have used $H_{\rm rad}=(\pi^2g_*/90)^{\frac{1}{2}} T^2/M_P$
(where $M_P \approx 2.4 \times 10^{18}$GeV). 
For the case of radiation domination ($p=0$) this gives 
$M_\Phi>T_{\rm dec} > 10^{16}\vert{\bf h}\vert^2$GeV, where we took
$g_*\sim 10^3$. Given that in these scenarios the asymmetry
is generated by, at the very least, the interference between a 
tree-level and one-loop diagram, it is always suppressed by some
small numbers times at least a square of the couplings ${\bf h}$, 
and often by higher powers of the couplings. Hence to produce a 
significant asymmetry one cannot have the coupling too small, and
conversely one needs the scalar field to have a mass not so far
below the GUT scale.

For the cosmologies we are primarily considering these bounds change very 
considerably. In {\it Model A\/} in which the Universe is dominated 
by a kinetic mode ($p=1$, or equivalently $n=6$) the 
constraint~(\ref{mass phi constr.}) relaxes to 
\begin{equation}
M_\Phi> T_{\rm dec} > 3 \times 10^6 {\rm GeV} \vert{\bf  h}\vert
      \left(\frac{T_{\rm reh}}{T_{\rm ns}}\right)^{\frac{1}{2}}
\,> \, 5 \, \vert{\bf  h}\vert \times 10^6 \,{\rm GeV}.       
\label{mass phi constr. 2}
\end{equation}
where we took $T_{\rm ns}=1$MeV and $g_*\sim 10^3$. 
%$ \vert{\bf  h}\vert \equiv ({\bf  h}{\bf  h}^\dag)^{1/2}$.
This should be compared with the energy scale $H_I$ which
characterizes this model at the beginning of the post-inflationary 
epoch. For a reheat temperature $T_{\rm ns}$ and a pure $n=6$
scaling after inflation  one finds \cite{jp} 
\begin{equation}
H_I \sim 10^7 \rm{GeV}
\left(\frac{T_{\rm reh}}{T_{\rm ns}}\right)^{\frac{1}{2}} .
\label{HI in kination}
\end{equation}
Thus the $M_\Phi$ can be sufficiently light that they are produced by
gravitational coupling in this mechanism along with all the other 
lighter ($m<H$) degrees of freedom. A little later, at a 
temperature $T \sim 0.1 H_I$ the strongly interacting degrees of
freedom begin to equilibrate (and define a real thermodynamic
temperature), while the $\Phi$ can decay without ever coming into
equilibrium. For smaller values of the coupling ($h\lsim 10^{-2}$) 
there may be some time for weak force mediated annihilation 
processes (with rate $\sim \alpha_w^2 T$) to act, and in this 
case the initial $e_R$ number density at the time of decay will 
be reduced somewhat relative to their initial value.

In the case of {\it Model B}, when the dominant component decays, we have
$p=2$ so that Eq.~(\ref{mass phi constr.}) gives an even milder bound on 
the mass of $\Phi$: 
\begin{equation}
 M_\Phi> T_{\rm dec} > 2\, \vert{\bf  h}\vert^{\frac{2}{3}}
\,\left(\frac{T_{\rm reh}}{T_{\rm ns}}\right)^{\frac{2}{3}} \, {\rm TeV}
\,> \, 3 \, \vert{\bf  h}\vert^{\frac{2}{3}} \, {\rm TeV}.
\label{mass phi constr. 3}
\end{equation}
In this case therefore the out-of-equilibrium
condition may in some cases (for sufficiently low $T_{\rm reh}$)
provide an even weaker
constraint on their masses than accelerator constraints 
from the flavour changing processes they can mediate.
More generally, it is certainly interesting to note 
that the mass scale is sufficiently low that models may be viable in which
the scalars are the supersymnmetric scalar partners of the standard model 
particles. We will return to this point in our conclusions.
Therefore in {\it models} of type {\it B\/} we can envisage the following
scenario. The universe attains a temperature $T\gg M_\Phi$
and the $M_\Phi$ are created by the fastest processes in similar 
quantities to the other degrees of freedom; as the temperature
falls they drop out of equilbrium and, when the temperature
$T_{\rm dec}$ is reached,  they decay. As in {\it Model A\/} one would 
need to consider carefully the different cases 
(depending on $\vert{\bf h}\vert$) in which the weak interactions 
can or cannot play a role in reducing the particle anti-particle
asymmetry in $\Phi$ before this decay occurs. 
One feature of (\ref{mass phi constr. 3}) should
immediately be noted, however, and we will return to it below:
The entropy release of these models which is of relevance 
in the present case is that which occurs between the time of
production of the $e_R$ asymmetry, $T_{\rm dec}$, and $T_{\rm reh}$.
From (\ref{mass phi constr. 3}) it follows that 
\begin{equation}
 \frac{T_{\rm dec}}{T_{\rm reh}} > 2\, \vert{\bf  h}\vert ^{\frac{2}{3}}
\,
\left(\frac{{\rm TeV}}{T_{\rm ns}^{{2}/{3}}
\;T_{\rm reh}^{1/3}}\right)
\, \gsim 10^5 \, \vert{\bf  h}\vert ^{\frac{2}{3}} ,
\label{dec-reh constraint}
\end{equation}
where the latter equality follows from (\ref{kin.eR.5}).
When it comes to producing a final baryon asymmetry this constraint 
will make it very difficult for models with any scaling much slower
than the kinetic mode ($\rho_\phi\propto a^{-6}$) 
to produce a reasonable final baryon
to entropy ratio. We will return to this below.

\subsection{Generation of the Asymmetry}

We now turn to the production of the asymmetry through the
out-of-equilibrium decay of these scalar fields. 
\begin{figure}[t]
\centering
\hspace*{-2mm}
\leavevmode\epsfxsize=14cm \epsfbox{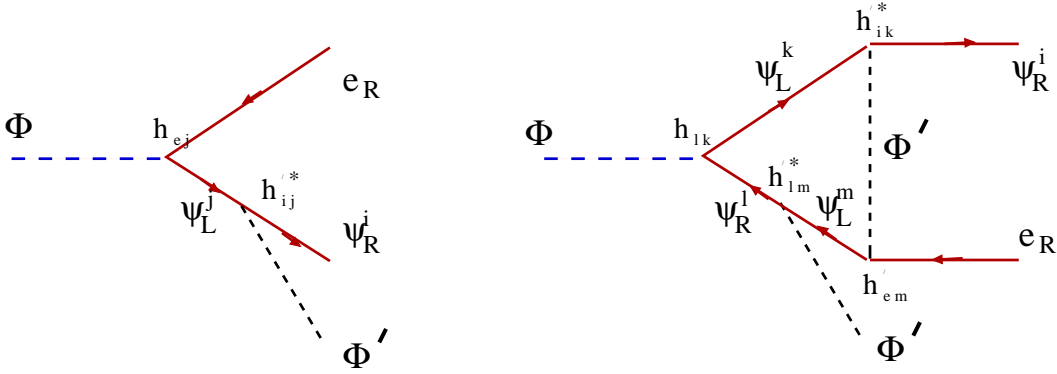}\\[3mm]
\caption[fig1]{\label{fig1} Tree and one loop diagrams for the
three body decay $\Phi \rightarrow \bar{e}_R \Psi^i_R \Phi'$,
with the appropriate couplings at the vertices. 
We assume that $\Phi$ is heavier than $\Phi'$. 
When the second outgoing lepton is a $\mu$ or $\tau$ lepton
the process produces net $e_R$ number.}
\end{figure}
As in any CP-violating out-of-equilibrium decay scenario one must go beyond
the tree-level decay and consider the interference between tree-level
and higher order diagrams to produce any net CP-violating effect. 
Further, since CPT theorem implies the equality between the total
decay rate for particles and anti-particles, we need at least two
channels containing different $e_R$ number in order to be able to
produce the asymmetry. It is to this end that we have taken the
$\Phi^a$ scalars to couple to more than one generation. Further,
to produce a CP-violating effect at one loop level we need at least two
scalars, just as one requires two heavy bosons in simple GUT scenarios 
({\it cf.} Ref.~\cite{kolbturner}). Here we do not try to be exhaustive 
in our consideration of the diagrams which give dominant contributions
in different parts of parameter space. In Figure 2 the 
two diagrams we consider here are shown, for the decay channel 
$\Phi \rightarrow \bar{e}_R \Psi^i_R \Phi'$ where the $\Phi'$
is assumed to be the lighter of the (at least) two scalars. 
Provided the second fermion $\Psi^i_R$  is of one of the 
two heavier lepton flavours, the process violates right
electron number. The rate of the 
corresponding anti-particle decay does not cancel if
the CP-violating interference terms between the two diagrams
have a pure complex part. Summing over the internal fermions
we have (see \cite{fukigita} for somewhat similar cases) that the
net CP-violating effect creating net $e_R$ number compared
to the tree-level decay is 
\begin{equation} 
\frac{\Gamma_{\Phi\rightarrow \bar{e}_R \Psi^i_R \Phi'} - 
\Gamma_{\bar{\Phi} \rightarrow e_R \bar{\Psi}^i_R \bar{\Phi}' }}
{\Gamma^{tot}_{\Phi} + \bar \Gamma^{tot}_{\bar{\Phi}}}
= \epsilon_p  \frac{ {\rm Im}\left[({\bf h} {\bf h'}^\dag)_{ei} 
( {\bf h'} {\bf h}^\dag {\bf h'} {\bf h'}^\dag)_{ie}  \right]}
    {{\rm Tr}\left[{\bf h}{\bf h}^\dag\right]} ,
\label{eR.4.1.a}
\end{equation}
where $\epsilon_p\sim 10^{-2}$ is the phase space factor
(and $i$ is summed over the non-electron indices). There is
also another pair of diagrams which differ only in that 
the $\Phi'$ emission on the external $e_R$ leg, which 
gives~(\ref{eR.4.1.a}) with the indices interchanged. 
From the result we see clearly that to obtain an effect at this order
we indeed need two scalars since, when ${\bf h}={\bf h'}$, the result in
Eq.~(\ref{eR.4.1.a}) vanishes. We note that a diagram in which the
$\Phi'$ on the external leg is the standard model Higgs 
could dominate if the ${\bf h'}$ couplings are all smaller 
than the Yukawa coupling of the $\tau$ lepton 
($y_\tau \approx 10^{-2}$).

We thus write the resultant $e_R$ asymmetry as 
\begin{equation}
\frac{e_R}{s}(T_{\rm dec}) \sim \frac{10^{-2}}{g_*}\,
\vert{\bf h}\vert^4 \, \delta_{CP} 
\label{eR.4.1.b}
\end{equation}
where $\delta_{CP}$ is proportional to the imaginary part
in (\ref{eR.4.1.a}). If all the couplings $h$ are of
the same order this can be of order one, while if there
is a hierarchy similar to that in the standard model,
it will be correspondingly smaller.

\subsection{The Baryon Asymmetry}

To arrive at the final baryon asymmetry the cases 
of {\it Model A\/} and {\it Model B\/} are quite different. In the former
case there is no entropy dilution and the final baryon to
entropy ratio is quite simply given by (\ref{eR.B2}), 
{\it i.e.} one third of the right electron to entropy ratio~(\ref{eR.4.1.b}).
We thus have 
\begin{equation}
\frac{n_B}{s} \sim \frac{10^{-2}}{g_*} \vert{\bf h}\vert^4 \delta_{CP},
\qquad (Model\; A) ,
\end{equation}
which for roughly equal on- and off-diagonal couplings in
${\bf h}$ and ${\bf h'}$ gives a baryon asymmetry of the required size
$n_B/s \sim 10^{-10}$  for couplings of the order $\sim 10^{-1} - 10^{-2}$. 
Note that this corresponds (from (\ref{mass phi constr. 2}))
to scalar masses as low as about $M_\phi \sim 100\,$TeV. 

In {\it Model B\/} there is the important entropy dilution factor, so
that for the final baryon asymmetry we find 
\begin{equation}
\frac{n_B}{s}\sim \frac{10^{-2}}{g_*}\,
\vert{\bf h}\vert^4 \, \delta_{CP} \,  
\left(\frac{T_{\rm reh}}{T_{\rm dec}}\right)^\frac{3(8-n)}{n},
\qquad (Model\; B)
 .
\label{eR.4.1.c}
\end{equation}
If we take the constraint given in (\ref{dec-reh constraint})
for couplings and $\delta_{CP}$ of order one, we can have a
baryon asymmetry compatible with that required for nucleosynthesis
only for a case $n \approx 6$, {\it i.e.} when the inflaton rolls in 
a kinetic energy dominated mode while it decays. The out-of-equilibrium 
decay condition for the $\Phi$ field~(\ref{mass phi constr. 3})
is in this case satisfied for $M_\phi$ as small as a few TeV. However
the constraints on the flavour changing neutral currents need to
be carefully considered, but can still be satisfied ({\it e.g.}
if one of the couplings $h_{e \mu}$ or $h_{e \tau}$ is
much smaller than the others). This provides in principle
an interesting probe at accelerators of these models in a
parameter range which is of interest.

For the standard matter scaling ($n=3$) \cite{dlr}, 
or indeed radiation scaling ($n=4$) during reheating,
the entropy dilution factor in (\ref{eR.4.1.c}) is much
too large to allow the generation of the required baryon
asymmetry, and the mechanism we have discussed is not
a viable one for baryogenesis in these cases.
In a different model it may be possible to relax the 
condition (\ref{eR.B1}) and reduce the dilution factor.
This would be appropriate for example if the $e_R$ continues
to be created all the way down to the electroweak scale, 
for example in a scenario in which the $\Phi$ particles
are themselves directly produced out of equilibrium by 
the inflaton decay all the way to that scale. Tuning the 
value of $T_{\rm reh}$ to be just enough to
keep the $e_R$ out of equilibrium until that time, {\it i.e.} to satisfy
(\ref{kin.eR.5}), the case $n=3$ gives a dilution by a factor 
$\sim 10^6$, and the case $n=3$ by $\sim 10^4$. A fairly copious
initial $e_R$ asymmetry must therefore be produced very close
to the electroweak scale in order to give the required baryon 
asymmetry. 

The model we have presented here is completely perturbative.
It is likely that there are models of non-perturbative decay
of a condensate of the $\Phi$ field in which the constraints 
inferred on the Yukawa coupling ${\bf h}$ may be relaxed.
One simple possibility would be  a variant of the well-known 
Affleck-Dine mechanism~\cite{ad}, with a scalar field $\Phi$  
charged under right-handed electron number, which oscillates 
and decays. This may occur for example in supersymmetric 
extensions of the Standard Model when an $A\Phi^3+h.c.$ term is present 
in the potential for a weakly coupled scalar field. When the field decays it  
creates a net right-handed electron number which is not suppressed 
by any coupling constant. The suppression may also potentially be
evaded without the scalar field being required to carry a 
right electron charge. This would occur if there were a resonant
decay into the electrons, which may in fact occur through precisely the
same Yukawa coupling discussed here. Because they do not
have the Yukawa coupling suppression, these mechanisms for producing
$e_R$ would leave more space for entropy release in models like
our model B. However, as we have pointed out, this will only work in
the special case that the $e_R$ is created very close to the electroweak
scale, and the temperature at which radiation domination begins
$T_{\rm reh}$ lies just far enough below the electroweak scale 
to keep the $e_R$ out of equilibrium at the electroweak scale.

\section{Anomaly and Magnetic Fields}

So far we have neglected the effects related to the abelian anomaly
discussed in the introduction. As described in \cite{js} the effect
of this term on a finite chemical potential $\mu_R$ for right electrons
is to cause an instability in modes with $k < \mu$. Naively this
instability can begin to grow when the corresponding mode enters the
horizon ({\it i.e.} $\mu \sim  H$) , but when the slowing effect of 
conductivity $\sigma$ in the plasma is taken into account the 
criterion becomes $\mu^2/\sigma \sim  H$. In the standard
case, in which the mode must be able to start to evolve before the
perturbative $e_R$ decay comes back into play, this corresponds to
the effect being important if $\mu/T \gsim 10^{-6}$. Here since the
expansion rate is such that the $e_R$ come back into equilibrium  
below the electroweak scale, the lower bound (which depends now 
simply on the expansion rate at the electroweak scale) will in 
fact be the same or larger. Moreover, while in the standard case 
the effect of the perturbative channel is important, and leads to 
the requirement that a significantly larger asymmetry than the
critical one be present initially in order for the source hypermagnetic 
field to survive to the electroweak scale, here the 
instability will simply develop all the way to that scale and there
will be no damping of the fields due to the appearance of the 
perturbative channel. 

The conclusions we can draw are as follows: 

$\bullet$ {\it Model A:\/} If the initial $e_R$ asymmetry is less than the 
critical value for hypermagnetic field generation, all our analysis 
given above holds. A baryon asymmetry
is produced of the same order, and thus for compatibility with the
observed asymmetry one must have the corresponding initial value
$\mu_R \sim 10^{-10}T$. For initial $\mu_R$ larger than the critical 
value, hypermagnetic field will be generated. The baryon asymmetry
generated will depend on the attained value of the
chemical potential $\mu_R$  (which reduces as the Chern-Simons
number grows in the condensate). However, the latter will always
be bounded below by the same critical value, and thus the baryon
asymmetry also (see also \cite{giovshap} for a discussion of the
full evolution of the dynamical equations). Thus we conclude 
that magnetic field generation from $\mu_R$ cannot be attained 
in this model with an altered
expansion rate, since it will always be associated with a baryon
asymmetry which is too large. On the other hand, the baryogenesis
mechanism from right electrons works perfectly well, and is
unaffected by the anomalous effects as the chemical potential is 
so small.

$\bullet$ {\it Model B:\/} The effects of the interplay of the baryon
generation and the instability causing the growth of
magnetic field are much more difficult to evaluate in this case,
and are in general model dependent. The fact that the entropy dilution
effect mandates a larger initial electron asymmetry, which would
then be subject to the instability, suggests that it may be
possible to find a model in which both magnetic field and the 
observed baryon asymmetry could be produced from the right electrons.
As was discussed above, the production of the baryon asymmetry would
require that a large $\mu_R$ chemical potential be produced very
close to the electroweak scale. The corollary is that, if
it is produced close to the electroweak scale, there will be little
time for the instability to evolve and create significant
seed fields. One way of getting around this would be in a 
model with a continuous sourcing of the $e_R$ asymmetry, so that
the contribution at earlier time may grow into a magnetic field.
The driving chemical potential is, however, itself being constantly
diluted by entropy production, and the energy in the resultant field 
also relative to the background energy density. To see whether seed 
fields of significant magnitude can survive to the electroweak
scale would require detailed study in a model of $e_R$ generation
quite different to that we have discussed.

\section{Conclusions}

We have considered here one aspect of cosmologies in which a scalar
field dominates the expansion rate prior to nucleosynthesis. 
Right-handed electrons may remain out of equilibrium until the
electroweak scale, so that  if they are generated the
$B+L$ processes of the standard model will lead to a non-zero
equilibrium density of baryons of the order of that in
the $e_R$. We have discussed two kinds of post-inflationary
cosmologies in which such a period of scalar field domination 
can occur: in the first the inflaton rolls away after inflation
into a kinetic mode in a steep potential, and the Universe is
`reheated' by the gravitational particle production at the end of
the inflationary epoch, while in the second the inflaton rolls
into a mode which can have a range of scalings and reheats
the Universe itself by decaying sufficiently slowly to give
a very low reheat temperature. We studied a specific model
for the generation of the right handed electron asymmetry
in which there are a set of scalars with CP-violating (and
flavour changing) couplings to the leptons. We showed that in
both scenarios such scalars can decay out of equilibrium at
quite low temperatures and produce the desired asymmetry.
While our models strongly favoured the case of kinetic mode
domination, which have little or no entropy release, we note that
in certain very special circumstances which may be satisfied in
other models the generation of the observed baryon asymmetry 
may still be possible in the standard reheating scenario
(with matter scaling during the reheating epoch). 
Finally we considered briefly the effect of the
abelian anomaly which destabilizes such charges, and concluded
that in the models with kinetic mode domination this effect is
unimportant for the baryon number generation, while in the case
with large entropy dilution it may be important and might allow 
the generation of magnetic field as in the case of standard radiation
domination.

Finally we return to the question of how this kind of mechanism
might be implemented in other particle physics models, in particular
in more popular (e.g. supersymmetric) extensions of the standard model.
In general one need not consider necessarily the generation of 
right-handed electron number, but the generation of 
{\it any CP odd charge which is effectively conserved
after its creation on a timescale which is longer than the expansion
rate of the Universe at the electroweak scale} (when the $B+L$ 
violating processes freeze-out). Given that the expansion rate 
at the electroweak scale can be enhanced in these models by many 
orders of magnitude -- up to a rate $\sim 10^{-11}T_{\rm ew}$ in models 
of type A, and  $\sim 10^{-6} T_{\rm ew}$ in models of type B -- scenarios
can be considered in which many of the lighter degrees of freedom
will drop out of equilibrium (for example the lighter right-handed
quarks). While in the standard model itself there would seem to be 
the obstacle of prohibitively small CP violation, in extensions 
there is generically new CP-violating structure in the added
sectors ({\it e.g.} in the chargino and squark mass matrices of the
minimal supersymmetric standard model). The problem of baryogenesis 
then becomes the problem of the generation prior to the electroweak 
scale of CP-odd approximately conserved charge using 
this structure. Given our observation that for very modest
masses (as low as a TeV for a particle with a coupling of order
one) the decay of these heavier particles occurs out of equilibrium 
in these cosmologies, there is clearly the interesting possibility
of sourcing CP-odd charges in this way, thus creating a baryon
asymmetry. We will treat these issues in detail in forthcoming work 
\cite{jp2}.

\section*{Acknowledgements}

We would like to thank Kimmo Kainulainen and Misha Shaposhnikov for useful
discussions. MJ thanks Sacha Davidson and Steve Abel
for discussion while this work was being completed.

%
%
%%%%%%%%%%%%%%%%%%%%%%%%%%%%%%%%%%%%%%%%%%%%%%%%%%%%%%%%%%%%%%%%%%%%%%%%%%%%%%%
%  REFERENCES
%%%%%%%%%%%%%%%%%%%%%%%%%%%%%%%%%%%%%%%%%%%%%%%%%%%%%%%%%%%%%%%%%%%%%%%%%%%%%%%
%
% These are "Nuclear Physics"-type macrros
% ----------------------------------------------------------------------------
%
\nc{\AP}[3]    {{\it Ann.\ Phys.\ }{{\bf #1}, {#2} {(#3)}}}
\nc{\NP}[3]    {{\it Nucl.\ Phys.\ } {{\bf #1}, {#2} {(#3)}}}
\nc{\PL}[3]    {{\it Phys.\ Lett.\ } {{\bf #1}, {#2} {(#3)}}}
\nc{\PR}[3]    {{\it Phys.\ Rev.\ } {{\bf #1}, {#2} {(#3)}}}
\nc{\PRL}[3]   {{\it Phys.\ Rev.\ Lett.\ } {{\bf #1}, {#2} {(#3)}}}
\nc{\PREP}[3]  {{\it Phys.\ Rep.\ }{{\bf #1}, {#2} {(#3)}}}
\nc{\RMP}[3]   {{\it Rev.\ Mod.\ Phys.\ }{{\bf #1}, {#2} {(#3)}}}
%\nc{\APJ}[3]   {{\it Astrophys.\ J.\ }{{\bf #1}, {#2} {(#3)}}}
\nc{\IBID}[3]  {{\it ibid.\ }{{\bf #1}, {#2} {(#3)}}}
%
% ----------------------------------------------------------------------------
%


\begin{thebibliography}{00}

\bibitem{Supernovae} 
  % DISCOVERY OF A SUPERNOVA EXPLOSION AT HALF THE AGE OF THE UNIVERSE 
  % AND ITS COSMOLOGICAL IMPLICATIONS.
  Supernova Cosmology Project (S. Perlmutter et al., {\it Nature}
   {\bf 391}, {51} (1998), astro-ph/9712212;
  % MEASUREMENTS OF OMEGA AND LAMBDA FROM 42 HIGH REDSHIFT SUPERNOVAE.
 S. Perlmutter et al.,
{\it Astrophys. J.} {\bf 516} (in press), astro-ph/9812133;
A.G. Riess {\it et. al.}, {\it Astron. J.} {\bf 116}, {1009} {(1998)}. 

\bibitem{quintessence}
See for example
%COSMIC CONCORDANCE AND QUINTESSENCE.
L. Wang, R.R. Caldwell, J.P. Ostriker and P.J. Steinhardt,
{\it Astrophys. J.} {\bf 530}, {17} {(2000)}, astro-ph/9901388;  
%NEW CONSTRAINTS FROM HIGH REDSHIFT SUPERNOVAE AND LENSING STATISTICS UPON
%SCALAR FIELD COSMOLOGIES.
I. Waga and J.A. Frieman, astro-ph/0001354. 

\bibitem{jf} 
% STRUCTURE FORMATION WITH A SELFTUNING SCALAR FIELD.
P. Ferreira and M. Joyce, \PRL{79}{4740}{1997}, astro-ph/9707286;
% COSMOLOGY WITH A PRIMORDIAL SCALING FIELD.
P. Ferreira and M. Joyce, \PR{D58}{023503}{1998}, astro-ph/9711102.

\bibitem{tracking}
%QUINTESSENCE, COSMIC COINCIDENCE, AND THE COSMOLOGICAL CONSTANT.
I. Zlatev, L. Wang and P.J. Steinhardt,
\PRL{82}{896}{1999},  astro-ph/9807002; P.J. Steinhardt, L. Wang and 
I. Zlatev, \PR{D59}{123504}{1999}, astro-ph/9812313. 

\bibitem{nucleosynthesis}
See for example P. Kernan and S. Sarkar, \PR{D54}{3681}{1996};
B.D. Fields, K. Kainulainen, K.A. Olive and D. Thomas,
New. Astron {\bf 1}, 77 (1996).

\bibitem{kolbturner}
E. Kolb and M.S. Turner, {\it The Early Universe}
(Frontiers in Physics, vol. 69),
Addison-Wesley Publishing Company (1990).

\bibitem{spokoiny} 
B. Spokoiny, \PL{B315}{40}{1993}.

\bibitem{jp} 
%TURNING AROUND THE SPHALERON BOUND: ELECTROWEAK BARYOGENESIS IN AN
% ALTERNATIVE POSTINFLATIONARY COSMOLOGY.
M. Joyce and T. Prokopec, \PR{D57}{6022}{1998}, hep-ph/9709320.

\bibitem{peeblesvilenkin}
%QUINTESSENTIAL INFLATION.
P.J.E. Peebles and A. Vilenkin, \PR{D59}{063505}{1999}, astro-ph/9810509.

\bibitem{mj} 
%ELECTROWEAK BARYOGENESIS AND THE EXPANSION RATE OF THE UNIVERSE.
M. Joyce, \PR{D55}{1875}{1997}, hep-ph/9606223.

\bibitem{Barrow} J.D. Barrow, \NP{B208}{501}{1982}.

\bibitem{KamTurner} M. Kamionkowski and M.S. Turner,
\PR{D42}{3310}{1990}; K. Griest, M. Kamionkowski and M.S. Turner,
\IBID{41}{3565}{1990}.

\bibitem{ford} 
%GRAVITATIONAL PARTICLE CREATION AND INFLATION.
L.H. Ford, \PR{D35}{2955}{1987}. 

\bibitem{dlr}  S. Davidson, M. Losada, and A. Riotto, hep-ph/0001301. 

\bibitem{tp}
  % A COMMENT ON BARYOGENESIS AT THE ELECTROWEAK SCALE
  % IN ALTERNATIVE COSMOLOGIES.
    T. Prokopec, hep-ph/0002181.

\bibitem{js} 
% PRIMORDIAL MAGNETIC FIELDS, RIGHTHANDED ELECTRONS, AND THE ABELIAN ANOMALY.
M. Joyce  and M.E. Shaposhnikov, \PRL{79}{1193}{1997},
      astro-ph/9703005. 

\bibitem{cdeo}
B. Campbell, S. Davidson, J. Ellis and K. Olive, \PL{297B}{118}{1992}.

\bibitem{cko} 
    % ON THE ERASURE AND REGENERATION OF THE PRIMORDIAL 
    % BARYON ASYMMETRY BY SPHALERONS.
   J.M. Cline, K. Kainulainen and K.A. Olive, \PRL{71}{2372}{1993},
      hep-ph/9304321; 
    % PROTECTING THE PRIMORDIAL BARYON ASYMMETRY FROM ERASURE BY SPHALERONS. 
   J.M. Cline, K. Kainulainen and K.A. Olive, \PR{D49}{6394}{1994},
               hep-ph/9401208.

\bibitem{giovshap}
%PRIMORDIAL MAGNETIC FIELDS, ANOMALOUS ISOCURVATURE FLUCTUATIONS AND BIG
%BANG NUCLEOSYNTHESIS.
M. Giovannini and M.E. Shaposhnikov, \PRL{80}{22}{1998}, hep-ph/9708303 and
%PRIMORDIAL HYPERMAGNETIC FIELDS AND TRIANGLE ANOMALY.
\PR{D57}{2186}{1998}, hep-ph/9710234. 

\bibitem{halliwellbarrow}
J.J. Halliwell, \PL{B185}{341}{1987}; J.D. Barrow, \PL{B187}{341}{1987}.
 
\bibitem{ratrapeeb}
B. Ratra and P.J.E. Peebles, \PR{D37}{3406}{1988}.

\bibitem{steinhardt}
%COSMOLOGICAL IMPRINT OF AN ENERGY COMPONENT WITH GENERAL EQUATION OF
%STATE.
R.R. Caldwell, R. Dave and P.J. Steinhardt, \PRL{80}{1582}{1998}, 
astro-ph/9708069. 

\bibitem{turner}
M.S. Turner, \PR{D28}{1243}{1983}.

\bibitem{dine} M. Dine, hep-th/0002047. 
  %TOWARDS A SOLUTION OF THE MODULI PROBLEMS OF STRING COSMOLOGY.

\bibitem{giovan}
%PRODUCTION AND DETECTION OF RELIC GRAVITONS 
%IN QUINTESSENTIAL INFLATIONARY MODELS.
M. Giovannini, \PR{D60}{123511}{1999}, astro-ph/9903004. 

\bibitem{rosati} 
% CAN THE INFLATON AND THE QUINTESSENCE SCALAR BE THE SAME FIELD?
Francesca Rosati,  Talk given at COSMO 99: 3rd International Conference 
on Particle Physics and the Early Universe, Trieste, Italy, 
27 Sep - 3 Oct 1999, hep-ph/0002090; 
%ON THE CONSTRUCTION OF QUINTESSENTIAL INFLATION MODELS.
M. Peloso and F. Rosati, {\it JHEP} {\bf 9912}, 026 (1999), hep-ph/9908271. 

\bibitem{preheating}
  % GUT BARYOGENESIS AFTER PREHEATING:
  %  NUMERICAL STUDY OF THE PRODUCTION AND DECAY OF X BOSONS.
E. W. Kolb, A. Riotto and I. I. Tkachev, \PL{B423}{348}{1998}, hep-ph/9801306. 

\bibitem{felderkofmanlinde}
%INFLATION AND PREHEATING IN NO MODELS.
G. Felder, L. Kofman and A. Linde, \PR{D60}{103505}{1999}, hep-ph/9903350.

\bibitem{gm} D. Bodeker, G. D. Moore, K. Rummukainen, \PR{D61}{056003}{2000},
             hep-ph/9907545; G. D. Moore,  hep-ph/0001216. 

\bibitem{jpt}
M. Joyce, T. Prokopec and N. Turok, \PR{D53}{2930}{1996}. 

\bibitem{leptogenesis}
% ELEMENTS OF BARYOGENESIS.
See for example W. Buchmuller and S. Fredenhagen, Presented 
at International School of
Astrophysics, Daniel Chalonge: 7th Course: Current Topics in
Astrofundamental Physics (A NATO Advanced Study Institute Euroconference),
Erice, Italy, 5-16 Dec 1999; hep-ph/0001098;
% MATTER ANTIMATTER ASYMMETRY AND NEUTRINO PROPERTIES.
W. Buchm\"uller and M. Pl\"umacher, contribution to the {\it Festschrift for
L.B. Okun}, submitted to {\it Phys. Rep.}, hep-ph/9904310. 

\bibitem{shaprub-review} 
See for example V. Rubakov and M.E. Shaposhnikov,   
{\it Phys.\ Usp.\ }{\bf 39}, 461 (1996).

\bibitem{sheryuan}
M. Sher and Y. Yuan, \PR{D44}{1461}{1991}.

\bibitem{sher}
%SCALAR MEDIATED FLAVOR CHANGING NEUTRAL CURRENTS.
M. Sher, in the {\it Proceedings of 29th International Conference 
on High-Energy Physics} (ICHEP 98), Vancouver, Canada, 23-29 Jul 1998,
hep-ph/9809590.

\bibitem{hall}
A. Antaramian, L.J. Hall and A. Ra\v sin, \PRL{69}{1871}{1992}. 

\bibitem{fukigita}  T. Yanagida and M. Yoshimura \PR{D23}{2048}{1981}; 
   M. Fukugita and T. Yanagida, \PL{B174}{45}{1986}. 

%\bibitem{felderkofmanlinde-grav}
%GRAVITATIONAL PARTICLE PRODUCTION AND THE MODULI PROBLEM.
%G. Felder, L. Kofman and A. Linde, hep-ph/9909508; 
%GRAVITINO PRODUCTION AFTER INFLATION.
%R. Kallosh, L. Kofman, A. Linde and A. Van Proeyen, hep-th/9907124; 


\bibitem{ad} I. Affleck and M. Dine, \NP{B249}{361}{1985};
%A NEW MECHANISM FOR BARYOGENESIS.
%BARYOGENESIS FROM FLAT DIRECTIONS OF THE SUPERSYMMETRIC STANDARD MODEL.
M. Dine, L. Randall and S. Thomas, \NP{B458}{291}{1996}, hep-ph/9507453.
% REGULATING THE BARYON ASYMMETRY IN NO SCALE AFFLECK-DINE BARYOGENESIS. 
B.A. Campbell, M.K. Gaillard, H. Murayama and K.A. Olive, 
\NP{B538}{351}{1999}, hep-ph/9805300. 


\bibitem{jp2} 
M. Joyce and T. Prokopec, in preparation.


\end{thebibliography}
\end{document}